\documentclass[twocolumn,showpacs,preprintnumbers,amsmath,amssymb]{revtex4}

\usepackage{graphicx} % Include figure files
\usepackage{dcolumn} % Align table columns on decimal point
\usepackage{bm} % bold math
\usepackage{Jonasmacros}

\begin{document}

\preprint{}

\title{Pauli Spin Blockade in Carbon Nanotube Double Quantum Dots}

\author{M.R. Buitelaar$^{1}$}
\author{J. Fransson$^{2}$}
\author{A.L. Cantone$^{1}$}
\author{C.G. Smith$^{1}$}
\author{D. Anderson$^{1}$}
\author{G.A.C. Jones$^{1}$}
\author{A. Ardavan$^{3}$}
\author{A.N. Khlobystov$^{4}$}
\author{A.A.R. Watt$^{4}$}
\author{K. Porfyrakis$^{4}$}
\author{G.A.D. Briggs$^{4}$}

\affiliation{$^{1}$ Cavendish Laboratory, University of Cambridge,
Cambridge, CB3 0HE, UK}

\affiliation{$^{2}$ Department of Physics and Materials Science,
Uppsala University, 751 21 Uppsala, Sweden}

\affiliation{$^{3}$ Clarendon Laboratory, Oxford University,
Oxford OX1 3PU, UK}

\affiliation{$^{4}$ Department of Materials, Oxford University,
Oxford OX1 3PH, UK}

\date{\today}

\begin{abstract}
We report Pauli spin blockade in an impurity defined carbon
nanotube double quantum dot. We observe a pronounced current
suppression for negative source-drain bias voltages which is
investigated for both symmetric and asymmetric coupling of the
quantum dots to the leads. The measured differential conductance
agrees well with a theoretical model of a double quantum dot
system in the spin-blockade regime which allows us to estimate the
occupation probabilities of the relevant singlet and triplet
states. This work shows that effective spin-to-charge conversion
in nanotube quantum dots is feasible and opens the possibility of
single-spin readout in a material that is not limited by hyperfine
interaction with nuclear spins.
\end{abstract}

\pacs{73.23.-b, 73.63.-b, 75.10.Jm, 85.35.-p}
% 73.63.Kv Quantum dots
% 73.23.Hk Coulomb blockade; single-electron tunneling
% 73.63.Fg Nanotubes
% 75.10.Jm Quantized spin models
% 85.35.Kt Nanotube devices
% 73.23.-b Electronic transport in mesoscopic systems
% 73.63.-b Electronic transport in nanoscale devices and structures
% 85.35.-p Nanoelectronic devices

%\keywords{Carbon nanotube, Quantum dots, Spin blockade}
%Use showkeys class option if keyword display desired
\maketitle

% ------------------------------------------------------------------
% Main text
% -------------------------------------------------------------------

\section{Introduction}

The electron spin is a natural two-level system and therefore
attractive as a quantum bit in quantum information processing
schemes. Spin qubits defined in quantum dots are of particular
interest because of the possibility to isolate, manipulate and
measure single spins \cite{Loss}. Much of the attraction of spin
qubits in quantum dots is also related to the relatively long time
over which a superposition of opposite spin states of a single
electron remains coherent. This long spin coherence time is a
direct result of the electron's small magnetic moment which
ensures it couples only weakly to its environment. The inevitable
problem this poses for the read-out of a single spin is elegantly
overcome in double quantum dot systems by converting the spin
information to a charge state using the phenomenon of Pauli spin
blockade which occurs when certain transitions between two quantum
dots are forbidden by spin selection rules \cite{Ono,Fransson} as
illustrated in Fig.~1. Significant experimental effort on spin
qubits defined in quantum dots in GaAs/AlGaAs heterostructures has
already resulted in the demonstration of driven coherent
oscillations of single electron spins \cite{Koppens1} and the
coherent exchange of two electron spins in a square-root-of-swap
quantum operation \cite{Petta}.

\begin{figure*}
\includegraphics[width=170mm]{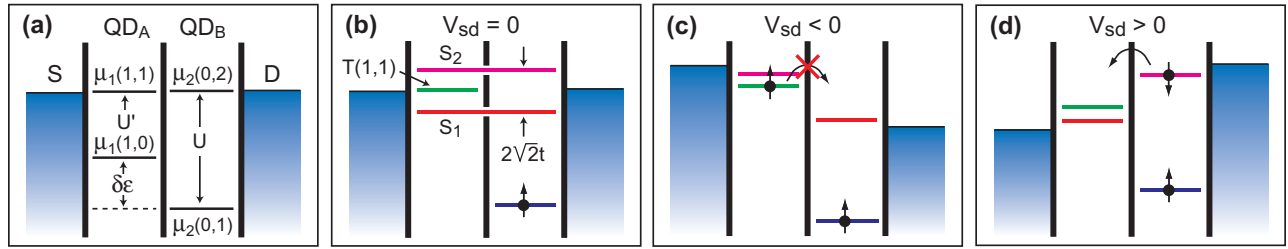}% size for 2column
\caption{\label{Fig1} (a) Schematics of the electrochemical
potentials of the relevant one and two-electron states of a double
quantum dot in the absence of a tunnel coupling between them. The
level offset between the single-particle states on both dots is
given by $\delta \epsilon$. The on-site charging energy and
electrostatic coupling energy are shown as $U$ and $U'$,
respectively. (b) When the tunnel coupling $t$ between the quantum
dots is significant, the S(1,1) and S(0,2) singlet states
hybridize to form molecular bonding (S$_1$) and anti-bonding
(S$_2$) singlet states which separate from the T(1,1) triplet
states. (c) When a negative bias voltage is applied and one of the
T(1,1) triplet states becomes occupied, the $(1,1) \rightarrow
(0,2)$ transition is not allowed because, by virtue of the Pauli
principle, the (0,2) state has to be a spin singlet and any
further current flow is blocked. (d) For opposite bias conditions
the $(0,2) \rightarrow (1,1)$ transition through the singlet
states is allowed. Note that for finite detuning, the S$_1$ and
S$_2$ singlet states (as well as the triplet states) are still
extended molecular states but are dominated by the (0,2) and (1,1)
charge states in the way indicated by the schematics.}
\end{figure*}

While significant as a proof-of-principle that single spin
manipulation and read-out in a solid-state environment is
feasible, these experiments also demonstrated that the spin
coherence time in these devices is limited by hyperfine
interactions with the Ga and As nuclei \cite{Koppens1, Petta,
Johnson1, Koppens2}. As a result, the number of coherent
single-spin rotations that can presently be observed within the
spin coherence time in GaAs based quantum dots is several orders
of magnitude below the typical figure-of-merit of $10^4$ quantum
operations for a fault tolerant quantum computer.

There is therefore a strong incentive to develop spin qubits in
materials in which hyperfine interactions are much reduced or
absent altogether. Carbon based materials such as carbon nanotubes
or graphene are excellent candidates in this respect. Due to the
absence of hyperfine coupling in the dominant $^{12}$C isotope,
the spin coherence times are expected to be very long
\cite{Semenov,Bulaev} while the recent observation of spin-orbit
interaction in nanotube quantum dots suggests the possibility of
electrical control of the spin states \cite{Bulaev, Ando, Huertas,
Kuemmeth}. In this work, we show that spin blockade is readily
observed in weakly coupled carbon nanotube quantum dots even for
many electrons on the nanotube and for temperatures of order one
Kelvin. We therefore conclude that effective spin-to-charge
conversion in carbon nanotube quantum dots is feasible and that
single-spin manipulation and readout in nanotubes is a promising
and realistic prospect.

\section{Nanotube devices}

The device we consider consists of an individual carbon nanotube
filled with Sc@C$_{82}$ molecules (that is a Sc atom inside a
C$_{82}$ cage) contacted by electron-beam defined palladium source
(S) and drain (D) electrodes that are separated by 300 nm. The
degenerately doped Si/SiO$_2$ substrate (300 nm oxide) is used as
back gate. The study of carbon nanotubes filled with Sc@C$_{82}$
is motivated by the long spin coherence times of the unpaired
spins on the encapsulated Sc atoms as observed by us in ensemble
measurements \cite{Morley}.

Measurements on twelve different devices suggest that even though
the presence of Sc@C$_{82}$ may lead to observable bandstructure
modification or charge transfer (doping) between the nanotubes and
Sc@C$_{82}$, the low-temperature transport properties are
remarkably similar to those of empty nanotubes \cite{regimes}.
These findings are consistent with recent transport experiments on
nanotubes with encapsulated C$_{60}$ molecules \cite{Utko, Quay}
as well as with recent density functional theory calculations of
spin interactions of chains of Sc@C$_{82}$ inside carbon nanotubes
\cite{Ge}. The conclusions of our present work will therefore
apply equally well to empty carbon nanotubes.

\section{Results and discussion}

The linear-response conductance of the device as a function of
gate voltage measured at temperature $T = 1.4$ K is shown in
Fig.~2(b) which displays a series of irregular conductance peaks.
A variation in peak height and spacing is common for carbon
nanotube quantum dots and is generally attributed to scattering by
defects along the nanotube. As demonstrated by a combination of
scanning probe and transport experiments \cite{Bockrath},
structural defects in particular lead to resonant electron
scattering in which the defects can be transparent or opaque
depending on the electrochemical potential (and hence the gate
voltage). Defects could be introduced in our nanotubes by, for
example, the acid treatment used to purify the samples from
magnetic impurities. However, defects are also commonly observed
in as-grown nanotubes such as those made by chemical vapor
deposition and for which typical scattering lengths of $\sim$ 150
nm were reported \cite{Bockrath}. Low-temperature transport
experiments on a number of our nanotube devices of the geometry
studied here indicate that most consist of a series of (two or
three) quantum dots.

Double quantum dots in carbon nanotubes in which the tunnel
barrier between the dots is due to a defect in the nanotube are
relatively common and have been studied previously in, e.g.,
Refs.~\cite{Mason, Jorgensen2}. Typically, the double quantum dot
is identified by the characteristic honeycomb pattern that
develops as two independent gate electrodes (each coupled to a
different quantum dot) are varied \cite{Wiel}. In our device we
only have the ability to vary a single gate electrode (the back
gate). Nevertheless, a detailed understanding of the physical
phenomena underlying the electronic transport properties can be
obtained by studying the differential conductance ($dI/dV$) as a
function of gate ($V_g$) and bias voltage ($V_{sd}$), as
demonstrated for double quantum dots by Ono \textit{et al}
\cite{Ono}. Here we focus on two pairs of peaks around $V_g = -
1.95$ V and $V_g = - 3.65$ V of which the the differential
conductance is shown in Figs.~2(c) and 2(d), respectively. The
most striking feature of the $dI/dV$ plots is the pronounced
asymmetry in the bias voltage and the current suppression and
appearance of negative differential conductance for negative
$V_{sd}$ (dark blue regions). Also note that while Fig.~2(c) is
approximately symmetric in gate voltage, this symmetry is clearly
broken in Fig.~2(d).

\begin{figure}
\includegraphics[width=84mm]{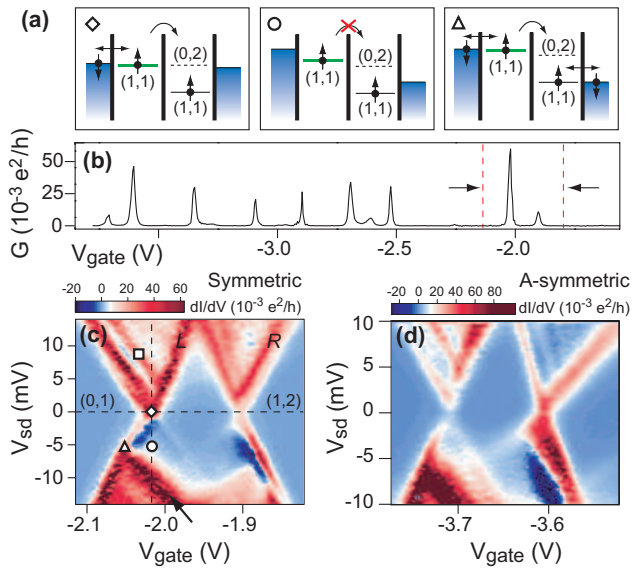}% size for 2column
\caption{\label{Fig2} (a) Schematic level diagrams for different
transport regimes corresponding to the measurements of panel (c),
see symbols. (b) Linear-response conductance of the device
measured at $T=1.4$ K. (c) Color-scale plot of the differential
conductance ($dI/dV$) as a function of source-drain bias voltage
($V_{sd}$) and gate voltage ($V_g$) for the gate region indicated
by the arrows in panel (b). Dark blue corresponds to negative
$dI/dV$. The ordered pairs ($n$,$m$) indicate the effective
electron occupancy in each quantum dot. (d) Differential
conductance for a different region of $V_g$, showing a pronounced
asymmetry in both bias and gate voltage.}
\end{figure}

The lack of periodicity in the linear-response conductance, the
pronounced negative differential conductance, as well as the
striking difference in the slopes of adjacent Coulomb diamonds
(most apparent in Fig.~2(d)) are clearly at odds with a model of a
single quantum dot. As we will show below, these features can be
explained well by invoking Pauli spin blockade in a double quantum
dot. The simplest model of two coupled single-level quantum dots
as introduced in Fig.~1 suffices to explain our measurements. This
might seem surprising given that the nanotubes will contain many
electrons but is justified by the large single-electron level
spacing $\Delta E$ (see below) and the simple even-odd shell
filling of carbon nanotube quantum dots \cite{Cobden1}. Recent
work on GaAs double dot systems containing more than two electrons
(up to $\sim$ 10) could also be explained in terms of effective
single-level quantum dots \cite{Liu, Johnson2}.

\subsection{Double quantum dot Hamiltonian}

To quantitatively compare the measurements in Fig.~2 with an
interpretation in terms of spin blockade, we have used a many-body
density matrix approach to calculate the current and population
numbers of the eigenstates of two coupled single-level quantum
dots. The double quantum dot (DQD) and the leads are modelled by
the Hamiltonian $\Hamil=\Hamil_L+\Hamil_R+\Hamil_{DQD}+\Hamil_T$,
where $\Hamil_{L(R)}$ models the left (right) lead in a free
electron like approximation. The double quantum dot is modelled by

\begin{eqnarray}
\Hamil_{DQD} & = & \sum_{n=A,B}\biggl(
    \sum_\sigma\dote{n\sigma}\ddagger{n\sigma}\dc{n\sigma}
    +Un_{n\up}n_{n\down}\biggr)
\nonumber\\&&
    {} +U'(n_{A\up}+n_{A\down})(n_{B\up}+n_{B\down})
\nonumber\\&&
    {} +\sum_\sigma(t\ddagger{A\sigma}\dc{B\sigma}+H.c.).
\label{eq-model}
\end{eqnarray}

In the model we have two levels, $\dote{A\sigma}$ and
$\dote{B\sigma}$, where $\sigma=\up,\down$ is the the electron
spin. Here, $\ddagger{A(B)\sigma}\ (\dc{A(B)\sigma})$ creates
(annihilates) an electron in quantum dot${A(B)}$ with spin
$\sigma$. The on-site Coulomb charging energy and electrostatic
coupling energy are denoted by $U$ and $U'$, respectively, whereas
$t$ is the tunneling rate between the quantum dots, see also
Fig.~1. The last term in $\Hamil$ accounts for the tunneling
between the leads and the DQD. This model neglects spin-orbit and
hyperfine interactions.

The Hamiltonian $\Hamil_{DQD}$ is transformed into diagonal form,
e.g. $\Hamil_{DQD}=\sum_{Nn}E_{Nn}\ket{N,n}\bra{N,n}$, where
$E_{Nn}$ is the energy for the eigenstate $\ket{N,n}$ with $N$
electrons, where $n$ is a state label (in our model there are 16
eigenstates, such that $N=0,\ n=1$, $N=1,\ n=1,\ldots,4$, $N=2,\
n=1,\ldots,6$, $N=3,\ n=1,\ldots,4$, and $N=4,\ n=1$). This
enables a many-body density matrix approach for calculating the
population number probabilities $P_{Nn}$ of the corresponding
state $\ket{N,n}$, to the first order approximation with respect
to the coupling $\Gamma^{L(R)}$ to the left (right) lead. This
order of approximation is based on that only the diagonal
transitions $\ket{N,n}\rightarrow\ket{N,n}$ are included, while
effects from off-diagonal transitions such as
$\ket{N,n}\rightarrow\ket{N,n'}$ would require a higher order
expansion of the rate equations for the population number
probabilities. Our calculation provides the dynamics of the
populations numbers as function of the bias voltage and
equilibrium electrochemical potential. Knowledge of the population
number probabilities enables calculation of the current and
differential conductance through the system, using standard
techniques. The approach provides complete knowledge of the matrix
elements $\bra{N,n}\Hamil_T\ket{N\pm1,m}$, for transitions between
states differing in electron number by one. This information
allows for detailed analysis of which states are involved in the
conductance and which state(s) are responsible for spin and
Coulomb blockade. The method is more thoroughly described in
Ref.~\cite{Fransson}.

\subsection{Double quantum dot model parameters}

We start by comparing the model predictions with the measurements
of Fig.~2(c). In the model, and in the analysis below, we assume
that the charging energy $U$ and the capacitive coupling to the
gate electrode is the same for both quantum dots. This
approximation is justified by the symmetry (in $V_g$) of the data
in Fig.~2(c). The result of the model calculations, using an
appropriate set of parameters which we will discuss below, is
shown in Fig.~3(a). The calculated differential conductance is in
good agreement with the experimental data of Fig.~2(c) for the low
bias regime while differences are observed at higher bias
voltages. Figure 4 illustrates how the differential conductance
(evaluated in Fig.~4 for the electrochemical potential
corresponding to the dotted vertical line in Fig.~3(a)) is related
to the occupation probabilities of the various one and
two-electron states. These calculations also show, see Fig.~4(b),
that the observed current suppression for negative $V_{sd}$ is
indeed the result of a near unity occupation probability of the
T(1,1) triplet states.

\begin{figure}
\includegraphics[width=80mm]{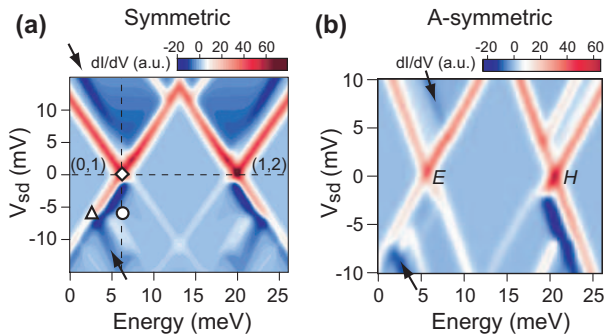}% size for 2column
\caption{\label{Fig3} (a) Color-scale plot of the calculated
differential conductance as a function of source-drain bias
voltage and gate voltage. The model parameters used are $\delta
\epsilon = 5$ meV, $U'=12$ meV, $U=17$ meV (both quantum dots) and
$t = 0.7$ meV. The relative voltage drop over the tunnel barrier
between the two quantum dots is $\Delta V_{QD} / V_{sd} \sim
0.15$. The temperature is set to $T=4$ K. The symbols correspond
to the schematics in Fig.~2. (b) Same as panel (a) for asymmetric
coupling of the double quantum dot to the leads as described in
the main text.}
\end{figure}

\begin{figure}[tr]
\includegraphics[width=73mm]{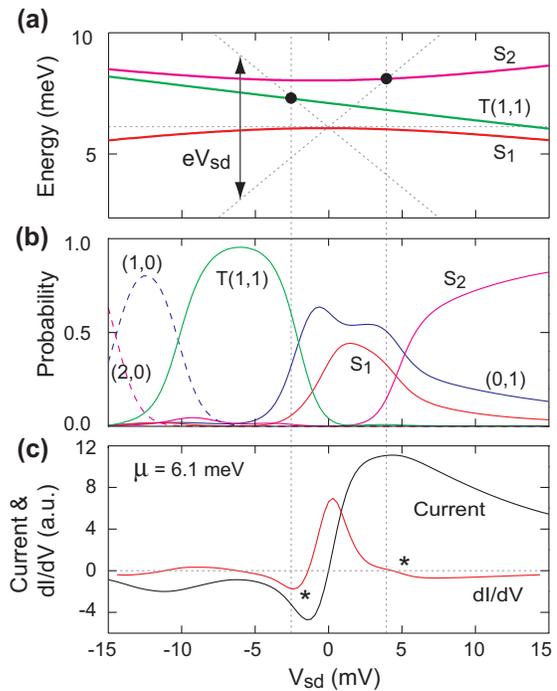}% size for 2column
\caption{\label{Fig4} (a) The current and $dI/dV$ can be
understood from the evolution of the two singlet states (S$_1$,
red and S$_2$, magenta) and the triplet states (green) as a
function of $V_{sd}$. The model parameters used here correspond to
the situation of Fig.~3(a). The diagonal lines indicate the
opening of the bias window as $V_{sd}$ is increased. The
electrochemical potential $\mu = 6.1$ meV corresponds to the
dotted vertical line in Fig.~3(a). (b) Calculated occupation
probabilities for the various one and two-electron states as a
function of $V_{sd}$. (c) Peaks in the current and $dI/dV$ are
observed when transitions between the one-electron and
two-electron states are energetically accessible.}
\end{figure}

The first model parameter to consider is the voltage drop $\Delta
V_{QD}$ over the tunnel barrier between the two quantum dots in
the presence of a source-drain bias voltage. The relative voltage
drop can be obtained from the slopes of the diamonds in the
experimental data, marked by $L$ and $R$ in Fig.~2(c) and from an
electrostatic model of the device \cite{Wiel,electrostatic} and
yields $\Delta V_{QD} / V_{sd} \sim 0.15$. As a result, the bias
voltage $V_{sd}$ acts as a knob that effectively controls the
level offset $\delta \epsilon$ (or `detuning') between the quantum
dots \cite{Wiel}. The magnitude of $\delta \epsilon$ can be
extracted from the excitation line indicated by the triangle in
Figs.~2(c) and 3(a). The position of the triangle in the $dI/dV$
plots corresponds to the situation in which (neglecting the tunnel
coupling) the electrochemical potentials on the left and right
quantum dot are aligned with the electrochemical potentials of the
source and drains electrodes, respectively, see right schematic in
Fig.~2(a). Taking into account the voltage drop between the
quantum dots, this yields $\delta \epsilon \sim 5$ meV. When the
bias voltage is increased beyond this point, the model calculation
shows, see Fig.~4(b), that the double dot gets trapped in a (1,0)
charge state (this excitation line is indicated by the lower arrow
in Fig.~3(a)). As a result, the current \textit{decreases} again
with increasing $V_{sd}$. In the experiment, however, this is not
observed. This difference can be understood considering that the
model does not account for inelastic scattering processes. This
approach is justified in the spin blockade regime since relaxation
due to e.g. electron-phonon interaction is strongly suppressed if
the transition involves a spin-flip \cite{Fujisawa}. On the other
hand, the (1,0) $\rightarrow$ (0,1) transition does not require a
spin-flip and energy relaxation by phonon emission will be
effective in the experiment. As a result, an \textit{increase} in
current is expected (and indeed observed) along this excitation
line, as indicated by the arrow in Fig.~2(c).

A similar reasoning would explain the absence of negative $dI/dV$
that is observed in the model for positive $V_{sd}$. For example,
the excitation line observed in the experiment (Fig.~2(c), square,
and upper arrow in Fig.~3(a)) could be due to a transition between
the S$_2$ and S$_1$ states which does not require a spin-flip
either. However, since the observed excitation line is exactly
parallel to the ground state transition, we tentatively attribute
it to a single-dot excitation, which yields $\Delta E \sim 4$ meV.
If we assume the conventional nanotube dispersion relation, the
level spacing is related to the nanotube length $L$ through
$\Delta E = h v_F / 4 L$, where $h$ is Planck's constant and $v_F
= 8.1 \times 10^5$ m/s is the Fermi velocity \cite{subband}. For
$\Delta E \sim 4$ meV this yields $L=200$ nm which, given the
source-drain separation of 300 nm, would imply two quantum dots of
similar size.

The electrostatic coupling energy $U'$ can be obtained from the
size of the main diamond (positive half) in Fig.~2(c) from which
we obtain $U' \sim 10-15$ meV \cite{Wiel}. The charging energy on
the individual quantum dots is more difficult to determine
exactly. The most satisfactory correspondence between the data and
model is obtained for $U \sim 17$ meV for both quantum dots,
consistent with dot lengths of order $\sim 100 - 200$ nm
\cite{Cobden1}. We verified that the conclusions of our work are
not sensitive to the exact value of $U$.

The ratio of the tunnel coupling to the level offset is crucial
for the observation of spin blockade and $t/\delta \epsilon \ll 1$
must be satisfied for it to be clearly observed \cite{Fransson}.
To appreciate this, consider that a current flow in the double dot
involves transitions between the one-electron state and the
two-electron singlet or triplet states. The one-electron state is
a superposition of the (0,1) and (1,0) charge states i.e. of the
kind $\alpha |0,1\rangle + \beta |1,0\rangle$ and the probability
of finding the electron in the energetically excited (1,0) state
depends directly on $t/\delta \epsilon$ \cite{Wiel}. Because in
the spin blockade regime, transferring an electron to the drain
electrode involves a transition from a triplet state to the (1,0)
charge state, a current flow will be strongly suppressed if $\beta
\ll \alpha$. Note that this is not the case for a current mediated
by the singlet states which follows the sequence
(1,1)$\rightarrow$ (0,2)$\rightarrow$ (0,1)$\rightarrow$ (1,1).

In our experiments, the tunnel coupling can be extracted from the
small leakage current of $\sim 100$ pA that is observed in the
spin blockade regime, see Fig.~5(a). As compared to the $\sim 5$
nA measured at positive $V_{sd}$ this implies a current
suppression of a factor of $\sim 50$. Given the previous estimate
for $\delta \epsilon$ of approximately 5 meV, the model requires
$t < 0.7$ meV to provide a similar suppression factor. The
observed leakage current also directly puts a lower bound of $e/I
\sim 2$ ns on the spin relaxation time $T_1$ in carbon nanotube
quantum dots. Note however that this is likely to be a strong
underestimate of the intrinsic spin-flip relaxation time in carbon
nanotubes and that the observed leakage current can be fully
accounted for by transitions mediated by the remaining finite
occupation probabilities of the one-electron and two-electron
singlet states. The strong dependence of the leakage current on
the tunnel coupling is illustrated in Fig.~5(b).

\begin{figure}
\includegraphics[width=82mm]{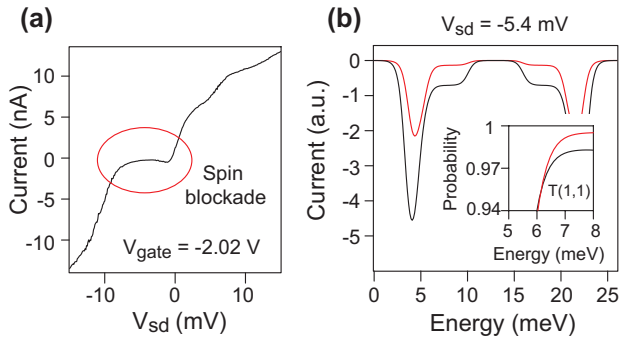}% size for 2column
\caption{\label{Fig5} (a) Measured current as a function of
$V_{sd}$ at gate voltage $V_g = -2.02$ V, corresponding to the
dotted vertical line in Fig.~2(c). The red oval indicated the
regime of Pauli spin blockade with measured leakage current of
$\sim 100$ pA. (b) Calculated current traces as a function of
electrochemical potential at $V_{sd} = -5.4$ mV for the model
parameters of Fig.~3(a). The tunnel couplings are $t=0.7$ meV
(black) and $t=0.3$ meV (red). Inset: maximum occupation
probability of the triplet states.}
\end{figure}

\subsection{Asymmetric coupling to the leads}

While the measurements of Fig.~3(a) are approximately symmetric in
gate voltage, this is not the case for the differential
conductance shown in Fig.~3(b). As compared to the symmetric
situation, there is a pronounced tilt in the slopes of the Coulomb
diamonds and strong negative differential conductance is only
observed along the right edge of the Coulomb diamond. In the model
these features are reproduced by introducing an asymmetry in the
potential drop at the source and drain electrodes and in the
tunnel couplings $\Gamma_s/\Gamma_d$ of the leads to the quantum
dots while keeping other parameters such as $U$, $U'$ and $\delta
\epsilon$ identical to the symmetric situation. The tunnel
coupling is set to $t=1.2$ meV. The result for $\Delta V_s/ \Delta
V_d = 3/2$ and $\Gamma_s/\Gamma_d = 1/5$ is shown in Fig.~3(b)
which corresponds well with the experimental data, see Fig.~2(d).

The dependence of the current in the Pauli spin blockade regime on
the asymmetry in the tunnel couplings for the model used here is
described in detail in Ref.~\cite{Fransson}. An intuitive way to
understand the effect of the asymmetry on the triplet occupation
probability (and therefore on the leakage current) is the
following: in the spin blockade regime as considered above, an
electron has a high probability to enter the double dot from the
source electrode to form a T(1,1) triplet state but a low
probability to exit to the drain (hence the high occupation
probability). An asymmetric coupling $\Gamma_s \ll \Gamma_d$, on
the other hand, has the precise opposite effect. The result is a
reduction in the triplet occupation probability such that negative
differential conductance is not observed along the edge of the
Coulomb diamond in the left part of Fig.~3(b).

The strong negative differential conductance along the Coulomb
diamond edge on the right side of Fig.~3(b) can be understood
considering the (broken) electron-hole symmetry of the double dot.
Whereas the linear-response conductance at point $E$ in Fig.~3(b)
corresponds to an electron moving from the source to drain
electrode, the conductance at point $H$ corresponds to a hole
moving in the opposite direction \cite{Wiel}. While the asymmetry
in the tunnel rates (partially) lifts the spin blockade for the
electron cycle, it enhances the blockade for the hole cycle (as it
moves in the opposite direction). The result is a strong T(1,1)
triplet occupation probability and negative conductance in the
$dI/dV$.

Note that several of the excitation lines in the model (indicated
by the arrows) are also observed in the data. The difference in
the polarity of the excitation (negative versus positive $dI/dV$
in the model and experiment, respectively) is attributed to energy
relaxation by e.g. phonon emission which is not accounted for in
the model. As discussed above, inelastic scattering will be
effective in the experiment for transitions that do not require a
spin-flip.

\section{Conclusions and outlook}

In conclusion, we present measurements of the differential
conductance of an impurity-defined carbon nanotube double quantum
dots showing Pauli spin blockade. The measurements are well
described by a theoretical model of the device which allows us the
estimate the relevant singlet and triplet occupation probabilities
of the double quantum dot. Since the phenomena of spin blockade
enables spin-to-charge conversion in quantum dots, our findings
present an important step towards single-spin read-out and spin
qubit operations in carbon-based devices that are not limited by
hyperfine interactions. An additional advantage of nanotubes in
this respect are the large energy scales observed here which
compare favorably to other systems such as lateral GaAs or Si
double quantum dots. In electron-spin resonance (ESR) experiments
\cite{Koppens1}, which are limited by photon assisted tunneling,
this would allow considerably larger oscillating fields and hence
much faster single-spin rotations. In fact, the significantly
larger g-factor of carbon nanotubes ($g \approx 2$) as compared to
GaAs devices ($g \approx -0.44$) would already provide a fivefold
gain.

For future experiments, control of the individual quantum dots and
the tunnel coupling is imperative. Fully tunable carbon nanotube
quantum dots have already been studied by a number of groups but
spin blockade had not been previously observed \cite{Mason,
Jorgensen2, Graber,Jorgensen,Sapmaz}. This seems surprising given
the clear signatures seen here at a relatively high temperature of
1.4 K. This difference is likely to be related to the condition
$t/\delta \epsilon \ll 1$ which must be satisfied for the
observation of spin blockade in double quantum dots. As our device
is rather small (dot sizes of order 100-200 nm) and does not make
use of metal top gates that would add to the overall device
capacitance, most quantum dot energy scales are nearly an order of
magnitude larger than in Refs. \cite{Mason, Jorgensen2,
Graber,Jorgensen}. We expect that when the device dimensions
and/or the tunnel couplings are sufficiently reduced, spin
blockade will also be observed in top gated carbon nanotube
quantum dots.

\section*{Acknowledgements}

We thank John Jefferson, John Morton and Karl Petersson for
discussions and Hisanori Shinohara for assistance with the
production of the fullerene material. GADB thanks EPSRC for a
Professional Research Fellowship (GR/S15808/01). This research is
part of QIP IRC (GR/S82176/01).

% End of Main text
%---------------------------------------------------------------------------

%-------------------------------------------------------------

\begin{thebibliography}{99}
\bibitem{regimes}
Depending on the coupling of the Sc@C$_{82}$ filled carbon
nanotubes to the leads, we have observed Coulomb blockade, Kondo
and Fano features.
\bibitem{electrostatic}
Assuming that the charging energies $U$ for the two quantum dots
are identical and that the gate capacitance ($C_g$) of the quantum
dots is small as compared to the capacitances of the quantum dots
to the source ($C_s$) and drain ($C_d$) electrodes as well as the
mutual capacitance ($C_m$), the relative voltage drop is given by
$\Delta V_{QD} / V_{sd} \sim (U-U')/(U+U')$. From the same model
we obtain $C_s = C_d \sim 4.9$ aF , $C_m \sim 13.3$ aF and $C_g
\sim 0.6$ aF.
\bibitem{subband}
Here we assume that the subband degeneracy is broken which will
generally be the case when atomic defects are present.
\end{thebibliography}
\end{document}